\input epsf
\documentstyle[12pt,epsf]{article}
\newlength{\pubnumber} 

\catcode`\@=11
\@addtoreset{equation}{section}

\def\section{\@startsection{section}{1}{\z@}{3.5ex plus 1ex minus .2ex}
 {2.3ex plus .2ex}{\large\bf}}
\def\subsection{\@startsection{subsection}{2}{\z@}{2.3ex plus .2ex}
 {2.3ex plus .2ex}{\bf}}

\font\it=cmti10 at 12pt
\font\ss=cmss10 at 12pt

 at 10truept

\ss 					%% default font
\begin{document}

\begin{titlepage}
\samepage{
\setcounter{page}{1}
\rightline{OUTP--02--02P}
\rightline{UNILE-CBR-2002-1}
\rightline{UFIFT--HEP--01--28}
\rightline{\tt hep-ph/yymmddd}
\rightline{January 2002}
\vfill
\begin{center}
 {\Large \bf  Seeking Experimental Probes of\\
                String Unification\footnote{To appear in the proceedings of 4th 
Meeting of the RTN Network and Workshop on Across the Present Energy Frontiers: 
Probing the Origin of Mass, Corfu, Greece, 10-13 Sep 2001.}}
\vfill
\vfill
 {\large Claudio Corian\`{o}$^{1,2}$\footnote{
        E-mail address: Claudio.Coriano@le.infn.it}
        $\,$and$\,$ Alon E. Faraggi$^{3}$\footnote{
        E-mail address: faraggi@thphys.ox.ac.uk}\\}
\vspace{.12in}

{\it $^1$Dipartimento di Fisica, Universita' di Lecce\\
 	I.N.F.N. Sezione di Lecce, Via Arnesano, 73100 Lecce, Italy\\}
\vspace{.075in}

{\it $^{2}$   Institute for Fundamental Theory, Department of Physics, \\
        University of Florida, Gainesville, FL 32611,USA\\}
\vspace{.075in}
{\it $^{3}$   Theoretical Physics Department, University of Oxford,\\
		Oxford OX1 3NP, Oxford, UK\\}
\end{center}
\vfill
\begin{abstract}
  {\rm
A general prediction of string unification is the existence of
exotic states with fractional charges under the free unbroken
Abelian generators of the underlying GUT symmetry. Such states
may be long--lived due to the existence of weakly broken gauge,
or local discrete, symmetries, and may serve as experimental
probes of string theory in forthcoming cosmic ray and dark
matter experiments.
}
\end{abstract}
\smallskip}
\end{titlepage}

\setcounter{footnote}{0}

% ========================= DEFINITIONS ===================================
\def\beq{\begin{equation}}
\def\eeq{\end{equation}}
\def\beqn{\begin{eqnarray}}
\def\eeqn{\end{eqnarray}}

\def\ie{{\it i.e.}}
\def\eg{{\it e.g.}}
\def\half{{\textstyle{1\over 2}}}
\def\third{{\textstyle {1\over3}}}
\def\quarter{{\textstyle {1\over4}}}
\def\m{{\tt -}}
\def\p{{\tt +}}

\def\slash#1{#1\hskip-6pt/\hskip6pt}
\def\slk{\slash{k}}
\def\GeV{\,{\rm GeV}}
\def\TeV{\,{\rm TeV}}
\def\y{\,{\rm y}}
\def\SM{Standard-Model }
\def\SUSY{supersymmetry }
\def\SSSM{supersymmetric standard model}
\def\vev#1{\left\langle #1\right\rangle}
\def\l{\langle}
\def\r{\rangle}

\def\Htw{{\tilde H}}
\def\chibar{{\overline{\chi}}}
\def\qbar{{\overline{q}}}
\def\ibar{{\overline{\imath}}}
\def\jbar{{\overline{\jmath}}}
\def\Hbar{{\overline{H}}}
\def\Qbar{{\overline{Q}}}
\def\abar{{\overline{a}}}
\def\alphabar{{\overline{\alpha}}}
\def\betabar{{\overline{\beta}}}
\def\tautwo{{ \tau_2 }}
\def\thetatwo{{ \vartheta_2 }}
\def\thetathree{{ \vartheta_3 }}
\def\thetafour{{ \vartheta_4 }}
\def\ttwo{{\vartheta_2}}
\def\tthree{{\vartheta_3}}
\def\tfour{{\vartheta_4}}
\def\ti{{\vartheta_i}}
\def\tj{{\vartheta_j}}
\def\tk{{\vartheta_k}}
\def\calF{{\cal F}}
\def\smallmatrix#1#2#3#4{{ {{#1}~{#2}\choose{#3}~{#4}} }}
\def\ab{{\alpha\beta}}
\def\Minv{{ (M^{-1}_\ab)_{ij} }}
\def\bone{{\bf 1}}
\def\ii{{(i)}}
\def\V{{\bf V}}
\def\b{{\bf b}}
\def\N{{\bf N}}
\def\t#1#2{{ \Theta\left\lbrack \matrix{ {#1}\cr {#2}\cr }\right\rbrack }}
\def\C#1#2{{ C\left\lbrack \matrix{ {#1}\cr {#2}\cr }\right\rbrack }}
\def\tp#1#2{{ \Theta'\left\lbrack \matrix{ {#1}\cr {#2}\cr }\right\rbrack }}
\def\tpp#1#2{{ \Theta''\left\lbrack \matrix{ {#1}\cr {#2}\cr }\right\rbrack }}
\def\l{\langle}
\def\r{\rangle}

%================== BLACKBOARD BOLD CHARACTERS ==============================

\def\inbar{\,\vrule height1.5ex width.4pt depth0pt}

\def\IC{\relax\hbox{$\inbar\kern-.3em{\rm C}$}}
\def\IQ{\relax\hbox{$\inbar\kern-.3em{\rm Q}$}}
\def\IR{\relax{\rm I\kern-.18em R}}
 \font\cmss=cmss10 \font\cmsss=cmss10 at 7pt
\def\IZ{\relax\ifmmode\mathchoice
 {\hbox{\cmss Z\kern-.4em Z}}{\hbox{\cmss Z\kern-.4em Z}}
 {\lower.9pt\hbox{\cmsss Z\kern-.4em Z}}
 {\lower1.2pt\hbox{\cmsss Z\kern-.4em Z}}\else{\cmss Z\kern-.4em Z}\fi}

%========================================================================
%          MACROS FOR REFERENCES
%========================================================================
\def\AEF{A.E. Faraggi}
\def\NPB#1#2#3{{\it Nucl.\ Phys.}\/ {\bf B#1} (#2) #3}
\def\PLB#1#2#3{{\it Phys.\ Lett.}\/ {\bf B#1} (#2) #3}
\def\PRD#1#2#3{{\it Phys.\ Rev.}\/ {\bf D#1} (#2) #3}
\def\PRL#1#2#3{{\it Phys.\ Rev.\ Lett.}\/ {\bf #1} (#2) #3}
\def\PRT#1#2#3{{\it Phys.\ Rep.}\/ {\bf#1} (#2) #3}
\def\MODA#1#2#3{{\it Mod.\ Phys.\ Lett.}\/ {\bf A#1} (#2) #3}
\def\IJMP#1#2#3{{\it Int.\ J.\ Mod.\ Phys.}\/ {\bf A#1} (#2) #3}
\def\nuvc#1#2#3{{\it Nuovo Cimento}\/ {\bf #1A} (#2) #3}
\def\RPP#1#2#3{{\it Rept.\ Prog.\ Phys.}\/ {\bf #1} (#2) #3}
\def\APJ#1#2#3{{\it Astrophys.\ J.}\/ {\bf #1} (#2) #3}
\def\APP#1#2#3{{\it Astropart.\ Phys.}\/ {\bf #1} (#2) #3}
\def\etal{{\it et al\/}}

%=============================================================================
\hyphenation{su-per-sym-met-ric non-su-per-sym-met-ric}
\hyphenation{space-time-super-sym-met-ric}
\hyphenation{mod-u-lar mod-u-lar--in-var-i-ant}
%=============================================================================

%============================== SECTION 1 ============================

\setcounter{footnote}{0}
String theory is the leading candidate for a theory of quantum
gravity. Although in itself an important achievement,
the primary challenge facing string theory is to prove
its relevance for experimental data. On the other front,
the Standard Particle Model successfully accounts for all
observations in contemporary accelerator and non--accelerator
experiments. However, despite this enormous success
the Standard Model is not satisfactory
as it leaves many issues unresolved, including:
the origin of the particle spectrum and interactions;
the experimental verification of the Higgs sector and its
incorporation in a fundamental theory;
finally the framework of point quantum field theories,
on which the Standard Model is based, is not compatible
with quantum gravity. Synthesizing these two fronts
is the domain of string phenomenology. Superstring phenomenology
serves the dual purpose of
developing the tools and methodology to confront
string theory with the experimental data, and
of providing the structures and framework
to try to understand how the building blocks of the
Standard Model may arise from a consistent theory of
quantum gravity. In this respect,
while much effort is being devoted to understanding
the structures of the Standard Model in the context
of various schemes beyond the Standard Model, these
attempts are in general deficient in the sense that they
assume additional structures for which there is no observational
need or evidence. String phenomenology on the other
hand has the advantage that the additional structures
are not added injudiciously, but are rather imposed by
the consistency of the theory. One should then make the most
strenuous effort to derive from string theory solely the
observed Standard Model physics. Once successful,
any left--over can then truly be considered as a
prediction of the theory, or of the specific string model.

In the past we discussed various possible signatures
of string theory, which included: specific patterns
of the supersymmetric spectrum \cite{dedes};
extra stringy $Z^\prime$'s \cite{zprime};
and stringy dark matter candidates \cite{sdm}. In this paper
we discuss the availability of string theory
candidates to explain the Ultra High Energy Cosmic Ray (UHECR)
events beyond the Greisen--Zatsepin--Kuzmin (GZK)
cutoff \cite{gzk}. In this respect one of the most fascinating
unexplained experimental observations is that of
Ultra High Energy Cosmic Rays with energies in excess of
the Greisen--Zatsepin--Kuzmin (GZK) bound \cite{uhecr}.
There are apparently no astrophysical sources
in the local neighborhood that can account for the
events. The shower profile of the
highest energy events is consistent with identification of the
primary particle as a hadron but not as a photon or a neutrino.
The ultrahigh energy events observed in the air shower arrays
have muonic composition indicative of hadrons.
The problem, however, is that the propagation of hadrons
over astrophysical distances is affected by the
existence of the cosmic background radiation, resulting
in the GZK cutoff on the maximum energy of cosmic ray
nucleons $E_{\rm GZK}\le10^{20}\;{\rm eV}$.
Similarly, photons of such high energies have a mean free path of less than
10Mpc due to scattering {}from the cosmic background radiation and
radio photons. Thus, unless the primary is a neutrino,
the sources must be nearby. On the other hand, the primary
cannot be a neutrino because the neutrino interacts very weakly
in the atmosphere. A neutrino primary would imply that the
depths of first scattering would be uniformly distributed
in column density, which is contrary to the observations.

One of the most intriguing possible solutions
is that the UHECR primaries originate {}from the decay of long--lived
super--heavy relics, with mass of the order of $10^{12-15}\;{\rm GeV}$
\cite{Berezinsky}.
In this case the primaries for the observed UHECR would originate
from decays in our galactic halo, and the GZK bound would not apply.
Furthermore, the profile of the primary UHECR indicates that
the heavy particle should decay into electrically charged
or strongly interacting particles.
{}From the particle physics perspective the meta--stable super--heavy
candidates should possess several properties.
First, there should exist a stabilization mechanism which produces
the super--heavy state with a lifetime of the order of
$10^{17}s\le \tau_X \le 10^{28}s$,
and still allows it to decay and account for the observed UHECR events.
Second, the required mass scale of the meta--stable state
should be of order $M_X~\sim~10^{12-13}{\rm GeV}$.
Finally, the abundance of the super--heavy relic
should satisfy the relation
$ ({\Omega_X/\Omega_{0}})({t_0/\tau_X})\sim5\times10^{-11}\;,$
to account for the observed flux of UHECR events.
Here $t_0$ is the age of the universe, $\tau_X$ the lifetime
of the meta--stable state, $\Omega_{0}$ is the critical mass density
and $\Omega_{X}$ is the relic mass density of the meta--stable state.

As we discuss here, superstring theory inherently possesses the ingredients
that naturally give rise to super--heavy meta--stable states.
The stabilization
mechanism arises in string theory due to the breaking of the non--Abelian
gauge symmetries by Wilson lines. This gives rise to states in the string
spectrum that carry fractional charges under the unbroken free
$U(1)$ generators
in the Cartan subalgebra of the original non--Abelian gauge symmetry.
The most apparent and well known such example is that of states
that carry fractional electric charge. However, as we discuss further below,
string models may also contain exotic states that carry the
standard Standard Model
charges, but carry fractional charge under an orthogonal free Abelian
subgroup of the original non--Abelian gauge symmetry. Such states
cannot fall into representations of the original unbroken non--Abelian
gauge group. Furthermore, they arise in string theory due to the
nontrivial topology of the string and the breaking of the non--Abelian
gauge symmetry by Wilson line. Thus, such states in general do not arise
in ordinary Grand Unified Theories, in which the non--Abelian gauge symmetries
are broken by the Higgs mechanism. The existence of fractionally
charged
states can be regarded as a general consequence of string unification,
or as a very specific string prediction. The question, however, is how
can such states reveal themselves in contemporary experiments.

The existence of fractionally charged states in string theory
obviously gives rise to a stabilization mechanism. The states
that carry fractional electric charge cannot decay due to electric
charge conservation. However, also those exotic states that carry standard
Standard Model charges but fractional $U(1)_{Z^\prime}$ charge
may be stable. This arises if the Standard Model states and the
Higgs multiplets
are identified with representations of the original GUT theory. In this
case even after the breaking of $U(1)_{Z^\prime}$ by a Higgs VEV there
remains
a discrete symmetry which forbids the decay of the exotic state
to the Standard Model states. In practice it is sufficient to demand that
vevs which break the discrete symmetry are sufficiently small.
The super--heavy
states can then decay via the nonrenormalizable operators
\beq
{{\langle V_1~~~\cdots~~~ V_N\rangle}\over{M_S^{N-3}}}~~~;~~~
M_S\sim10^{17-18}{\rm GeV}
\label{nonreno}
\eeq
which are produced from exchange of heavy string modes.
The lifetime of the meta--stable relic is then given by
\beq
\tau_X~\approx~ {1\over{M_X}}\left(
{{M_S}\over{M_X}}\right)^{2(N-3)}
\label{taux}
\eeq
Additionally, string theory may naturally produce mass scales of the
required order, $M_X\approx10^{12-13}{\rm GeV}$. Such mass scales
arise due to the existence of an hidden sector which typically contains
non--Abelian $SU(n)$ or $SO(2n)$ group factors. Thus, the mass scale
of the hidden gauge groups is fixed by the hidden sector gauge
dynamics. Therefore, in the same way that the color $SU(3)_C$
hadronic dynamics are fixed by the boundary conditions at
the Planck scale and the $SU(3)_C$ matter content, the
hidden hadron dynamics are set by the same initial conditions
and by the hidden sector gauge and matter content,
$M_X~\sim~\Lambda_{\rm hidden}^{\alpha_s,M_S}(N,n_f)$.
Finally, the fact that $M_X\sim10^{12-13}{\rm GeV}$ implies
that the super--heavy relic is not produced in thermal
equilibrium and some other production mechanism is responsible
for generating the abundance of super--heavy relic. This may
arise from gravitational production \cite{ckr} or from inflaton
decay following a period of inflation.

\begin{figure}[t]
\centerline{\epsfxsize 3.0 truein \epsfbox {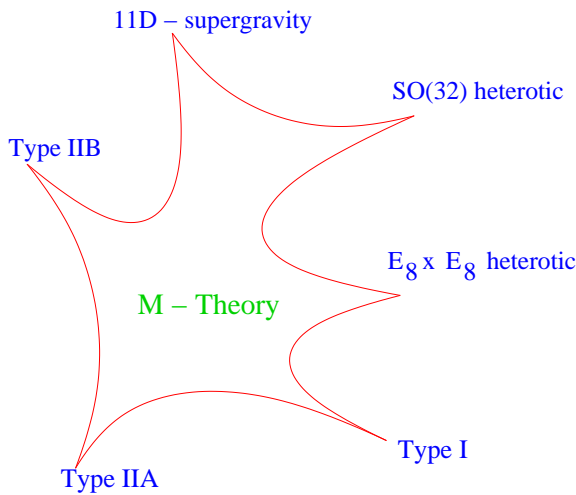}}
\caption{M--theory picture of string theory}
\label{mtheory}
\end{figure}

Fractionally charged states are generic in the perturbative heterotic string
theory. We may however question whether this will remain a general consequence
of string theory also in the nonperturbative regime. Naturally a definite
answer to this question is not possible at present due to the non--existence
of a nonpurterbative formulation of string theory. However, we may try
to contemplate the understanding of string theory that emerges
from string dualities. In this picture, as depicted in fig.
(\ref{mtheory}) the different string
theories, including eleven dimensional supergravity are
perturbative limits of a more fundamental theory, dubbed M--theory.
In this context, we may question as well the utility of any
of the perturbative string limits to try to capture our
experimental reality. One may put the bar a bit higher
and question the use of string/M theory in the first place.
Clearly, the fact that string/M theory provides a self--consistent
framework for quantum gravity does not yet imply its relevance
for physical reality, and indeed alternative approaches do
exist \cite{ashtekar}. We may furthermore infer from the approach
of ref. \cite{fm} to quantum mechanics that basic particle properties
may fundamentally be seen to arise differently from the conventional
Hilbert space constructions, which underly all of the string theories.
Thus, it may eventually be revealed that the Hilbert space
construction, and with it the notion of a particle with definite
properties, is an effective description, rather than a fundamental
one. With these issues in mind, then what sense is there in seeking
experimental probes of string theory, in general, and in a specific
perturbative limit, in particular. To formulate a logical sequence
we must go back and examine the premise of the Standard Model.
%%%%%%%%%%%%%%%%%%%%%%INSERT FIGURE HERE%%%%%%%%%%%%%%%%%%%%%%%%%%%%%%%%%
%%%%%%%%%%%%%%%%%%%%%END OF FIGURE%%%%%%%%%%%%%%%%%%%%%%%%%%%%%%%%%%%%%%%%
The Standard Particle Model is the objective reality
as we perceive it in contemporary experiments. The
technological and sociological complexity of the collider
experiments by which the Standard Model was revealed represent
the pinnacle of scientific achievement. The Standard Particle
Model is composed of three group factors
$SU(3)_C\times SU(2)_L\times U(1)_Y$, three generations
of chiral matter states that carry charges under the
three group factors, and a Higgs sector. The Higgs sector
has not yet been discovered experimentally and its
elucidation is perhaps the most burning question in
basic physics. However, examining the matter sector
alone we note that, in the process of its experimental
discovery, the gauge quantum numbers are in fact free experimental
parameters. A naive count therefore gives $3\cdot3\cdot6=54$
free parameters, counting the 3 group factors, three generations,
and the 6 $\{Q,L,D,U,E,N\}$ states in each generation, including the
right--handed neutrino. Now, a true miracle of the Standard Particle
Model is the fact that each generation fits into a single
representation of $SO(10)$. From the parameter account this miracle
entails the reduction of the number of free gauge quantum parameters
from 54 to 3! A true miracle indeed! One may therefore take the
view that the essence of the Standard Model is its embedding in
a Grand Unified Theory! The GUT framework, however, provides only
partial unification, and cannot explain the origins of flavor,
hierarchy and gravity, which are clearly additional features
of the objective reality. String theory provides the only
currently available self--consistent theoretical framework
in which these additional structures are unified. Therefore,
at present string theory provides the only available tool at our
disposal to explore the unification of flavor, hierarchy and gravity.

None of the perturbative string theories is more fundamental than the
others. This conclusion is clear from the mere fact that the
perturbative string theories are perturbative, and do not follow
from a fundamental physical principle. The utility of the
different string theories is precisely as the perturbation theory
of the more fundamental structure, which we may call M--theory.
Hypothesizing that the true vacuum of the world does indeed
lie somewhere in the region enclosed in fig. (\ref{mtheory}),
the different perturbative limits probe
different properties of the true vacuum. From our objective
reality, as perceived by the Standard Model, we may hypothesize
that a fundamental feature of the true vacuum is the embedding
of the Standard Model multiplets in $SO(10)$ representations.
The only perturbative limit which enables this embedding is
the perturbative heterotic string limit. This indicates
that if we choose to preserve the $SO(10)$ GUT embedding of
the Standard Model spectrum then the perturbative limit that
we should use is the heterotic string limit, whereas the other
perturbative limits may be more useful to learn about other
properties of the true vacuum. In this limit the $SO(10)$
symmetry is broken by Wilson line directly at the string level,
rather than by a Higgs VEV in the effective low energy field
theory. The string spectrum then will necessarily contain
exotic states which fractional charges under the unbroken
free $U(1)$'s in the Cartan sub--algebra of $SO(10)$.
In this effective perturbative limit of the
true vacuum such states are realized and are endemic.
The existence and self--consistency of the perturbative description
itself then indicates that such states are realized and may have
experimental manifestation. Following the lead suggested
in ref. \cite{fm}, and therefore keeping in mind that the
notion of a particle with definite properties is in any case
an effective description, how and whether such states
will appear in the nonpurterbative limit is immaterial.

The requirement that a realistic string vacuum admits
the $SO(10)$ embedding of the three chiral generations
is highly restrictive. The reason being that
string vacua typically contain additional $U(1)$ generators,
beyond those present in the Standard Model or its
GUT extensions. This facilitates finding combinations
of the $U(1)$ currents which reproduce the correct
Standard Model hypercharge assignments.
The expense is that the $SO(10)$ embedding and the
canonical normalization of the weak hypercharge is
lost. Consequently there exist many models that
do not admit the $SO(10)$ embedding \cite{nonso10},
whereas those that possess the $SO(10)$ embedding are
less abundant.
A class of string models that possess the $SO(10)$ embedding
are those constructed in the free fermionic formulation.
These models correspond to $Z_2\times Z_2$ orbifold compactification
at the free fermionic point in the Narain moduli space,
augmented with Wilson lines that break the $SO(10)$
symmetry. The models themselves have been built in the free fermionic
language \cite{FFF}, but can be translated to orbifold language.
The general structure of these models has been amply reviewed
in the past and we refer interested readers to the original
literature \cite{revamp}
and the reviews for the details \cite{reviews}.
Here we recap the main structure which is relevant for our
discussion of the meta--stable superheavy string relics.

The models can be seen to be constructed in two stages. The
first stage corresponds to the so--called NAHE set of boundary
condition basis vectors, typically denoted by
$\{{\bf1},S,b_1,b_2,b_3,X\}$ \cite{reviews},
and corresponds to the $Z_2\times Z_2$
orbifold compactification \cite{ztwo}. The three twisted sectors of the
$Z_2\times Z_2$ orbifold produce in these models the
three Standard Model chiral generations. The untwisted
sector produced the gravity and gauge multiplets and, plus
one additional sector, produces also the Standard Model Higgs
multiplets. In addition the orbifold spectrum contains hidden sector
matter states that transform in the vectorial representation
of the hidden $SO(16)$ subgroup. This hidden matter arises in
the string models due to the breaking pattern
$E_8\times E_8~\rightarrow~SO(16)\times SO(16)$ by
a GSO projection, which also breaks
$E_6~\rightarrow~SO(10)\times U(1)$. At the level of the NAHE
set the observable GUT symmetry is $SO(10)$,
with 24 chiral super--generations in the chiral 16 representation
of $SO(10)$.

The second stage consists of adding
to the NAHE set three additional boundary condition basis vectors,
typically denoted by $\{\alpha,\beta,\gamma\}$,
which correspond Wilson lines in the orbifold language. These
additional basis vectors break the $SO(10)$ symmetry to one
of its subgroups, where the cases of $SU(5)\times U(1)$, $SO(6)\times SO(4)$
or $SU(3)\times SU(2)\times U(1)^2$ have lead to quasi--realistic models.
At the same time the additional basis vectors reduce the number
of twisted chiral generations to three generations,
one from each of the twisted sectors $b_1$, $b_2$ and $b_3$.
The important point to emphasize from the discussion thus far
is that all the states which are identified with the Standard
Model states arise from the orbifold sectors. All these states are
$SO(10)$ representations, which are reduced
into representations of the final unbroken $SO(10)$
subgroup. In this construction, therefore, the Standard
Model admits the $SO(10)$ embedding, and the weak--hypercharge has
the canonical $SO(10)$ normalization.

The realistic free fermionic models provide an arena
in which many of the phenomenological issues pertaining to the Standard Model
as well as those pertaining to supersymmetric unification can be examined
from the view point of perturbative quantum gravity. We refer interested
readers
to the previous review articles and references therein \cite{reviews}. The
issues
studied include: top quark mass prediction; fermion masses and mixing; proton
stability and neutrino masses; gauge coupling unification; squark degeneracy;
derivation of string models
with solely the MSSM spectrum in the low energy effective field theory. These
achievements demonstrate the utility of the free fermionic models as a
laboratory
to study these phenomenological issues in the context of a potentially
fundamental
theory. Alternatively the models provide the means of connecting string theory
with experimental data.

We now turn to the discussion of the exotic states as
a possible experimental probe of string theory.
In addition to the ``standard'' spectrum from the orbifold sectors,
there exist in the heterotic--string models ``exotic'' spectrum which
cannot fit into $SO(10)$ multiplets. This spectrum arises from
sectors which contain the Wilson line breaking sectors, and produces
the exotic matter in vector--like representations.
Their interaction terms in the
superpotential are obtained by calculating the
correlators between vertex operators.
The non--vanishing correlators must be invariant under all the
symmetries and the string selection rules.
The exotic ``Wilsonian'' matter states appear in the free fermionic
models in vector--like representations, and obtain mass terms from
cubic level or nonrenormalizable terms in the superpotential.
In general, unlike the ``standard'' spectrum, the ``exotic'' spectrum is
highly model dependent. We can however classify the exotic matter
according the patterns of the $SO(10)$ symmetry breaking by the specific
sectors. The $SU(5)\times U(1)$ and $SO(6)\times SO(4)$
type sectors produce states with electric charges $\pm1/2$.
Similar to QCD the fractional charges may be confined by a hidden sector
gauge group \cite{elnfc}. The resulting integrally charged bound
states then produce meta--stable superheavy matter. Similar to QCD
the mass scale of the bound states is fixed by the initial conditions
at the unification scale, and by the gauge and matter content of the
confining gauge groups. Mass scales of required order of $10^{12-13}{\rm GeV}$
appear very naturally in realistic heterotic string models due to the
existence of the hidden sector.
The fractionally charged
constituents are stable due to electric charge conservation, and the bound
states may decay through the nonrenormalizable terms (\ref{nonreno}).
Depending on the order $N$, the lifetime from Eq. (\ref{taux})
may be in the appropriate range to account for the flux of observed
UHECR events above the GZK cutoff \cite{elnfc}. However, in addition to
the lightest neutral bound states there exist in this model
also long lived meta--stable charged bound states, whose abundance
is comparable to that of the neutral states \cite{cfp}.
Constraints on the abundance of stable charged heavy matter
then places an additional constraint on the lifetime of
this form of UHECR candidates \cite{cfp}.

In addition to the fractionally charged states, the free fermionic
standard--like models contain states, which arise from
$SU(3)\times SU(2)\times U(1)^2$ type sectors, and
carry the regular charges under the Standard Model,
but carry ``fractional'' charges under the $U(1)_{Z^\prime}\in SO(10)$
symmetry. These states can be color
triplets, electroweak doublets, or Standard Model singlets
and may be good dark matter candidates \cite{sdm}.
The meta--stability of this type of states arises because
of their fractional $U(1)_{Z^\prime}$ charge. Namely, the fact that
the Standard Model states possess the $SO(10)$ embedding,
implies that there exist a discrete symmetry which protects
the exotic matter from decaying into the lighter Standard Model states.
We must additionally insure that the $U(1)_{Z^\prime}$ symmetry breaking
VEVs, break the discrete symmetry sufficiently
weakly. The uniton is such a color triplet that has been motivated
to exist at an intermediate energy scale due to its possible
role in facilitating heterotic--string gauge coupling unification.
It forms bound states with ordinary down and up quarks. The mass
of the uniton is generated from nonrenormalizable terms and
can be of order $10^{12-13}{\rm GeV}$, as required to explain
the UHECR events. Additionally, if the uniton is to contribute
substantially to the dark matter, the lightest bound state
must be neutral and the heavier charges states must be unstable.
However, contrary to the case of the fractionally charged states,
in uniton charged bound states can decay through $W^\pm$ radiation of
the ordinary quark with which it binds. Lastly, the free fermionic
Standard--like models contains Standard Model singlets that carry
fractional $U(1)_{Z^\prime}\in SO(10)$ charge. Such states
may be semi--stable provided that the discrete symmetry is
broken sufficiently weakly. Moreover, similar to the states
with fractional electric charge, they may transform under a
hidden sector non--Abelian gauge group and may their
mass scale may therefore be fixed by the confining
hidden sector scale. Being neutral, they provide ideal
dark matter and UHECR candidates.

Superstring models provide a variety of candidates with differing
properties that may account for the observed UHECR events.
The phenomenological challenge is to develop
the tools that will discern between the different candidates,
by confronting their intrinsic properties
with the observed spectrum of the cosmic ray showers.
The UHECR data, however, opens up new probes to the GUT
and string scale physics. The point is that in the analysis
of the decay products of the meta--stable states one must
extrapolate measured parameters from the low scale, at which
they are measured, to the high--scale of the hypothesized meta--stable
state. In this extrapolation, which covers more than 10 orders of magnitude
in energy scales, one must make some judicious assumptions
in regard to the particle content. Thus one may hope that the
extrapolation itself will enable to differentiate between
different assumptions in regard to the physics in the extrapolation range.
This methodology is very similar to that employed successfully in the
case of gauge coupling unification in supersymmetric versus
non--supersymmetric cases. There the motivation for the extrapolation
arises from the hypothesis of unification and one can show that it is
consistent only if one includes the supersymmetric spectrum.
Similarly, in the case of the of the UHECR events, the motivation for
the high scale are the events themselves and the possibility
to explain them with the super--heavy meta--stable matter. The
extrapolated parameters are the QCD fragmentation functions and
similarly one must include in the evolution whatever physics
is assumed to exist in the desert. In ref. \cite{cf}
supersymmetric fragmentation functions were developed for this
purpose. Furthermore, such functions may also be used in the
analysis of the cosmic rays showers, which arise from the
collision of the primaries with the atmosphere nuclei at a center of mass
energies of order $\sim100{\rm TeV}$. Most exciting, however, is perhaps
the fact that the forthcoming Pierre Auger and EUSO experiments
will explore precisely the physics of the UHECR above the GZK cutoff!
The hypothesized meta--stable super--heavy string relics
may then serve as experimental probes of the string physics,
provided that we are able to develop the phenomenological tools
to decipher their predicted properties, such as their fractional
electric or $U(1)_{Z^\prime}$ charge!

C.C. thanks the IFT Group at the Univ. of Florida at Gainesville
for hospitality and for partial financial support. A.F. thanks
the PPARC for financial support.
%=========================================================================
%======================== REFERENCES =====================================
%=========================================================================

\bibliographystyle{unsrt}

\vfill\eject
\end{document}